\documentclass[prb,floatfix,showpacs,twocolumn]{revtex4-1}
\usepackage{amsmath,amssymb,graphicx}

\begin{document}
\title{Continuous replica-symmetry breaking in mean-field spin-glass  models: Perturbation expansion without the replica trick}

 \author{V.  Jani\v{s}, A. Kauch, and A. Kl\'\i\v c}

\affiliation{Institute of Physics, Academy of Sciences of the Czech
Republic, Na Slovance 2, CZ-18221 Praha, Czech Republic }
\email{janis@fzu.cz}

\date{\today}

\begin{abstract}
The full mean-field solution of spin glass models with a continuous order-parameter function is not directly available and approximate schemes must be used to assess its properties. The averaged physical quantities are to be represented via the replica trick and the limit to zero number of replicas is to be performed for each of them. To avoid this we introduce a perturbation expansion for a mean-field free-energy functional with a continuous order-parameter function without the need to refer to the replica trick. The expansion can be used to calculate all physical quantities in all mean-field spin-glass models and at all temperatures, including zero temperature. The small expansion parameter is a difference between the continuous order-parameter function and the corresponding order parameter from the solution with one level of replica-symmetry breaking. The first correction beyond the approximation with one level of replica-symmetry breaking is explicitly evaluated in the glassy phase of the Sherrington-Kirkpatrick model.        
\end{abstract}

\maketitle

\section{Introduction}
 \label{sec:Intro}

Strongly disordered and frustrated spin systems, the prominent examples of which are spin glasses,  display a complicated low-temperature behavior. Due to frustration the standard homogeneous long-range orders are precluded and replaced by orderings with peculiar properties. First of all, the low-temperature phase shows everywhere  ergodicity breaking that cannot be removed or circumvented by the application of external fields or measurable sources.\cite{Palmer82} The free energy of such systems manifests in low temperatures a complex landscape with almost degenerate metastable states the description of which demands advanced mathematical tools.\cite{Mezard87} The usual way to handle the low-temperature behavior of spin glasses and other frustrated systems is the replica trick transforming static, quenched averaging to a dynamical, annealed one.\cite{Dotsenko01} The limit to zero of the replication index (number of replicas) in physical quantities  does not, however,  work as a simple perturbation (loop) expansion as initially introduced. One has to use non-perturbative techniques to reach a thermodynamically consistent solution in the limit of zero replicas. Parisi found a way how to break the symmetry of the order parameters in the replicated space of a mean-field, Sherrington-Kirkpatrick model.\cite{Parisi80} The dependence of the thermodynamic equilibrium state of spin glasses on the replication index and the structure of the order parameters in the replicated phase space reflects volatility of the thermodynamic limit to boundary conditions or the initial state used for the equilibration process. This is demonstrated by the concept of real replicas used to reconstruct homogeneity of the free energy in mean-field spin-glass models.\cite{Janis05c}
  
A thermodynamically consistent mean-field solution of spin glass models with the full, continuous replica-symmetry breaking (RSB) cannot be reached directly but only via either an iterative scheme or a perturbation expansion. The most direct way to approach the solution with a continuous replica-symmetry breaking is to use discrete hierarchies of mathematical replicas. It is, however, almost impossible to go quantitatively beyond a two-level (2RSB) solution.\cite{Parisi80} Moreover, approximations with finite numbers of replica hierarchies replace the continuous order-parameter function by a set of delta functions from which one cannot deduce a detailed structure of the distribution of the equilibrium order parameters. Alternatively, one can expand the full solution near the critical transition point to the spin-glass phase.\cite{Janis06b,Janis08a,Crisanti10a}  Presently, the most advanced construction of the solution with continuous RSB is a high-order perturbation expansion of a solution of the Parisi nonlinear differential equation resolved numerically by means of a pseudo-spectral code and Pad\'e approximants.\cite{Crisanti02} These expansions  are applicable only to continuous transitions and to temperatures not too far below the critical point.                  
 
Only a few attempts have been made to determine the structure of the solution with full continuous RSB at very low temperatures of the Sherrington-Kirkpatrick model. Iterative solutions with a high number of replica hierarchies were proposed either with the aid of the renormalization group\cite{Oppermann05} or  an expansion around the spherical model,\cite{Crisanti10b} or the replica-symmetric solution.\cite{Crisanti11} They all try to assess the impact of the continuous order-parameter function and make conclusions on low-temperature properties of the Sherrington-Kirkpatrick model.  These approaches rely on the replica trick and a smooth transition from the replica-symmetric to the full RSB solution.     

There are generalized spin-glass models with a structure of the phase space of the order parameters differing from that of the Ising spin glass. The mean-field solutions of the Potts glass,\cite{Elderfield83} p-spin glass,\cite{Gardner85} or quadrupolar glass\cite{Goldbart85b} show intervals of temperature  where a first step toward the Parisi solution in the replica trick, the so-called one-level replica symmetry breaking (1RSB), is locally stable and for some parameters the transition to the glassy phase is discontinuous. Except for an asymptotic expansion near the continuous transition to the glassy phase of the Potts model,\cite{Janis11} there are no other approaches that could describe the coexistence and transitions between the 1RSB and continuous RSB solutions in these models, in particular at low temperatures.   

The aim of this paper is to develop an approximate scheme that would be able to interpolate between the 1RSB and the full continuous RSB solutions in general mean-field spin-glass models. We construct it so that to be applicable in the whole low-temperature glassy phase without referring to the replica trick and covering both continuous and discontinuous transitions. The starting point will be an explicit representation for the free energy with a continuous order-parameter function from Ref.~\onlinecite{Janis08b} where we rescale its variables so that the zero-temperature limit is easily accessible. To allow for the limit to zero temperature we will have to peel off the part corresponding to the 1RSB solution from the full free energy. We obtain a suitable form of the free energy that will allow us to introduce a perturbation expansion around the 1RSB solution in powers of the continuous order-parameter function, or better of its correction to the respective order parameter from the 1RSB solution. We evaluate the expansion explicitly in the first order for the Ising spin glass without external magnetic field and assess its reliability by comparing its results with the high-order expansion around the critical temperature to the glassy phase of Ref.~[\onlinecite{Crisanti02}] and with Monte-Carlo simulations.  

We show in Sec.~\ref{sec:FE} how a free-energy functional containing the 1RSB order parameters together with a continuous order-parameter function  can be derived from the limit of free energies with finite-many hierarchies of replica generations breaking the replica symmetry.  We introduce the appropriate scaling so that the limit to zero temperature can explicitly be performed. Stationarity equations for the limiting free energy without replicas  are derived in Sec.~\ref{sec:SEQ}. The idea and relations needed to construct a perturbation expansion around the 1RSB solution in powers of the continuous order-parameter function are presented in Sec.~\ref{sec:PE}.  The lowest order approximation in closed form is derived in Sec.~\ref{sec:FOA}. Numerical results from the first-order of the perturbation expansion are presented and compared with other methods in Sec.~\ref{sec:Results}. We summarize the salient properties of the approximate construction in the final Section~\ref{sec:Conclusions}.      

\section{Generating free-energy functional}
 \label{sec:FE}
 
We cannot avoid introducing replicas and hierarchies of the replicated phase space if we want to describe thermodynamic equilibrium of glassy systems. Independently whether we do it via the replica trick and the replica symmetry breaking scheme or via a successive use of real replicas of the phase space to enforce thermodynamic homogeneity, we end up with the same result when the replication index is  analytically continued  to real numbers. The result is a series of free energy densities labelled by a number $K$ of replica hierarchies used. Free energy with $K$ replica hierarchies uses $2K + 1$ order parameters. We have   $\Delta \chi_{l}, m_{l}$ with $l= 1,\ldots, K$, where $\Delta \chi_{l}$ stands for the overlap susceptibility between the original spins and those from the $l$th hierarchical level and $m_{l}$ is a replication index connected with the $l$th replica hierarchy.  The remaining parameter $q$ is the  averaged square of the local magnetization after $K$ replications. This free energy  
can be represented explicitly as\cite{Janis05c}
\begin{subequations}\label{eq:mf}
 \begin{widetext} 
  \begin{multline} \label{eq:mf-avfe} f^K(q,\Delta\chi_1,\ldots,\Delta
    \chi_K; m_1,\ldots,m_K) = - \frac 1\beta \ln 2 %\\ 
    + \frac \beta 4
    \sum_{l=1}^K m_l\Delta\chi_l\left[2\left(q
        +  \sum_{i=l+1}^{K}\Delta\chi_{i}\right) + \Delta\chi_l\right] \\
    -\frac\beta 4 \left(1-q -\sum_{l=1}^K \Delta\chi_l\right)^2  -
    \frac 1\beta \int_{-\infty}^{\infty} \mathcal{D}\eta\ \ln Z_K\ ,
  \end{multline}
  \end{widetext}
  where we used a sequence of partition functions \pagebreak[1]
  \begin{equation}\label{eq:mf-hierarchy}
    Z_{l} =  \left[\int_{-\infty}^{\infty}\mathcal{D}\lambda_l\
      Z_{l-1}^{m_l}\right]^{1/m_l}\ .
  \end{equation}
\end{subequations}
The initial partition sum for the Ising spin glass reads\\ $Z_0 = \cosh\left[\beta\left(h
    + \eta\sqrt{q} + \sum_{l=1}^{K}\lambda_l \sqrt{\Delta\chi_l}
  \right)\right]$. It is the partition sum of the original spin model affected by the interaction with the replicated spins represented via fluctuating Gaussian fields $\lambda_{1},\ldots,\lambda_{K}$.  We denoted the Gaussian normalized differential
$\mathcal{D}\lambda \equiv {\rm d}\lambda\ e^{-\lambda^2/2}/\sqrt{2\pi}$.

The physical interpretation of free energy $f^{K}$ as discussed in Ref.~\onlinecite{Janis08b} is as follows. We take the partition sum of the original model $Z_{0}$, replicate it $m_{1}$ times and average over the spins from the first replica hierarchy. We downscale the result by a power $1/m_{1}$ to keep the correct spin normalization. The new effective partition function $Z_{1}$ is replicated $m_{2}$ times and averaged over the spins from the second level of the replicated spins. After downscaling we go on with higher replica levels. If a partition sum $Z_{l}$ is thermodynamically homogeneous it becomes independent of the scaling parameter $m_{l+1}$. This  just happens if $\Delta\chi_{l+1}=0$. When not,  we have to go to a higher level of replica-symmetry breaking. Stationarity conditions then only minimize deviations from the global homogeneity and make free energy $f_{K}$ at least locally homogenous, that is, with respect to infinitesimal variations of the replication indices.  The full RSB solution with a continuous order-parameter function is then obtained in the limit $K\to\infty$ with $\Delta\chi_{l}\sim 1/K$.      

Before we proceed to the continuous limit, $K\to\infty$, we rescale the order parameters so that to simplify the explicit temperature dependence and to allow for a straightforward limit to low and eventually zero temperature. The explicit limit to zero temperature can be performed if we rescale the replication indices $m_{l}\to \mu_{l}= \beta m_{l}$ and set apart the first overlap susceptibility and the replication index that we denote $\chi_{0}, \mu_{0}$, respectively. The rescaled hierarchical free free energy is     
\begin{widetext}
 \begin{multline} \label{eq:muf-discrete} f^{K+1}(q_{0},\chi_{0}, \mu_{0}; \Delta\chi_1,\ldots,\Delta     \chi_K;\mu_1,\ldots,\mu_K) \\ = - \frac 1\beta \ln 2 -\frac\beta 4 \left(1-q_{0} - \chi_{0}- \sum_{l=1}^K \Delta\chi_l\right)^2  + \frac 14  \mu_{0}\chi_{0}\left[2\left(q_{0}
        +  \sum_{l=1}^{K}\Delta\chi_{l}\right) + \chi_{0}\right]     \\ +
    \frac 14 \sum_{l=1}^K \mu_l\Delta\chi_l\left[2\left(q_{0}
        +  \sum_{i=l+1}^{K}\Delta\chi_{i}\right) + \Delta\chi_l\right]
 - \frac 1{\mu_{K}}\left\langle \ln \left[\left\langle \ldots \left\langle Z_{0}(h_{\eta} + \lambda\sqrt{\chi_{0}} + \Lambda_{K})^{\mu_{0}/\beta}\right\rangle_{\lambda}^{\mu_{1}/\mu_{0}}\ldots \right\rangle_{\lambda_{K}}^{\mu_{K}/\mu_{K- 1}}\right] \right \rangle_{\eta} \ ,
  \end{multline}
\end{widetext}
where we denoted $h_\eta \equiv h + \eta\sqrt{q_{0}}$, $\Lambda_{K}= \sum_{l=1}^{K}\lambda_l \sqrt{\Delta\chi_l}$, $\langle X(\lambda_{l})\rangle_{\lambda_{l}} = \int_{-\infty}^\infty \mathcal{D}\lambda_{l} X(\lambda_{l})$, and relabeled $q\to q_{0}$. 

The asymptotic expansion near the critical temperature in  Ref.~\onlinecite{Janis06b,Janis08a} proves that in the continuous limit $K\to\infty$ we indeed have $\Delta\chi_{l}= O(K^{-1})$ and $\Delta \mu_{l}= \mu_{l} - \mu_{l+1}= O(K^{-1})$. We denote $X= K^{-1}\sum_{l=1}^{K}\Delta \chi_{l}$ and $x= K^{-1}\sum_{l=1}^{n}\Delta \chi_{l}$ for $x \in [n/K, (n+1)/K]$. Then the overlap susceptibilities $\Delta\chi_{l}$ are no longer order parameters in the continuous limit. They only form an index set for the continuously distributed replication indices $\mu(x)$. The generating free-energy functional for the solution with continuous RSB can be represented as a functional\cite{Janis08b} 
\begin{multline}\label{eq:FE-continuous} f(q_{0},\chi_{0}, \mu_{0}; X, \mu(x)) = - \frac 1\beta \ln 2  - \frac  \beta4 (1 - q_{0} - \chi_{0} - X)^2 \\ + \frac 14  \mu_{0}\chi_{0}\left[2\left(q_{0} + X\right) + \chi_{0}\right]  + \frac 12 \int_{0}^{X}dx \mu(x) \left[q_{0} + X - x\right] \\
- \left\langle g_{\mu}(X, h_{\eta})\right\rangle_{\eta}\ . 
\end{multline} %
The interacting part of the energy $g_{\mu}(x,h)$ obeys a Parisi-like non-linear differential equation
\begin{align}\label{eq:g0-DE}
  \frac{\partial g_{\mu}(x,h)}{\partial x} &= \frac {1}{2}
  \left[\frac{\partial^2 g_{\mu}(x,h)}{\partial h^2} + \mu(x)
    \left(\frac{\partial g_{\mu}(x,h)} {\partial h} \right)^2 \right]
\end{align}
as can be determined from the limit $K\to\infty$  in the same manner as in Ref.~\onlinecite{Duplantier81}. Its solution can conveniently be represented as\cite{Janis08b} 
% \begin{subequations}\label{eq:g}
\begin{multline}\label{eq:g0}
  g_{\mu}(x,h) = \mathbb E_0(h;x,0)\circ g_{\mu}(h)  \\ \equiv
  \mathbb T_{y} \exp\left\{\frac 12 \int_0^x d y
    \left[\partial_{\bar{h}}^2 \right.\right. \\ 
  \left.\left. + \ \mu(y) g_{\mu}^{\prime}(y;h + \bar{h})\partial_{\bar{h}}
    \right] \right . \bigg\} g_{\mu}(h + \bar{h})\bigg|_{\bar{h}=0} \ ,
\end{multline}
where we used prime to denote the derivative with respect to the magnetic field
$h$, $g_{\mu}^{\prime}(y, h)\equiv \partial_h g_{\mu}(y,h)$ and introduced a
"time-ordering" operator $\mathbb T_{y}$ ordering products of
$y$-dependent non-commuting operators from left to right in the
$y$-decreasing succession. The time-ordered exponential is then 
defined as a power series of multiple integrals
\begin{multline*} \mathbb T_{y} \exp\left\{\int_{a}^{b} d y
    \widehat{O}(y)\right\} \equiv 1 \\ + \sum_{n=1}^\infty
  \int_{a}^{b} d y_1
  \int_a^{y_1}d y_2\ldots\int_0^{y_{n-1}}\!\! d
  y_n \widehat{O}(y_1)\ldots \widehat{O}(y_n)\ .
\end{multline*} Time-ordering operators are a standard tool in
representing quantum many-body perturbation expansions.

The initial interacting free energy being propagated by the evolution operator $\mathbb E_0(h;X,0)$ is the interacting part of the 1RSB free energy 
\begin{multline}\label{eq:g_{mu}-Zero}
g_{\mu}(h) = \frac 1\mu_{0} \ln \int_{-\infty}^{\infty}\frac {d \phi}{\sqrt{2\pi}}\\   e^{-\phi^{2}/2} \left[ \cosh\left(\beta(h + \phi\sqrt{\chi_{0}})\right) \right]^{\mu_{0}/\beta}\ .
\end{multline}

To complete the expression for the free energy with continuous RSB we have to add an equation for function $g_{\mu}^{\prime}(x,h)$.  From the definition of the evolution operator $\mathbb E_0$ we obtain
directly
\begin{subequations}\label{eq:gx-prime}
\begin{multline}\label{eq:gx1}
  \frac {\partial g_{\mu}(x,h)}{\partial h} = \mathbb
  E_0(h;x,0)\circ g_{\mu}^\prime(h) \\ + \frac 12 \int_0^x
  d y\ \mu(y)\mathbb E_0(h;x,y)\circ \left[g_{\mu}^{\prime} (y,
    h)\partial_{h} g_{\mu}^{\prime} (y, h)\right]\ .
\end{multline}
The solution to this integral equation can be represented via the fundamental 
evolution operator for this theory with a shifted  $\mathbb T$-ordered exponential
\begin{align}\label{eq:E-equation}
  g_{\mu}^{\prime}(x,h) &= \mathbb E(h;x,0)\circ g_{\mu}^\prime(h) \nonumber \\
  &\equiv \mathbb T_y \exp\left\{ \int_0^x d y
    \left[\frac 12 \partial_{\bar{h}}^2 \right.\right. \nonumber \\
  &\quad \left.\left.  + \mu(y) g_{\mu}^{\prime}(x,h +
      \bar{h})\partial_{\bar{h}} \right] \right\} g_{\mu}^{\prime}(h +
  \bar{h})\bigg|_{\bar{h}=0}\ .
\end{align}
 \end{subequations}

The equilibrium state is a stationary solution of free energy $f(q_{0},\chi_{0}, \mu_{0}; X, \mu(x))$ from Eq.~\eqref{eq:FE-continuous}. It is invariant with respect to infinitesimal variations of the scalar order parameters $q_{0},\chi_{0},\mu_{0}, X$ and the continuous order-parameter function $\mu(x)$ for $x \in [0,X]$. We can relate our variables  with the standard Parisi representation of the continuous order-parameter function $q_{P}(x_{P})$  via a transformation  $x_{P} \to \beta^{-1}[\mu_{0} + \mu(X - x)]$ and  $q_{P}(x_{P}) \to q_{0} + \chi_{0} + x$. 

\section {Stationarity equations}
\label{sec:SEQ}

The advantage of the representation of the free energy with a continuous RSB from Eqs.~\eqref{eq:FE-continuous} - \eqref{eq:gx-prime} is that we can derive the stationarity equations in the standard way used in statistical mechanics. Evaluating first the derivatives with respect to $q_{0}$, $\chi_{0}$ and $\mu_{0}$ we obtain
\begin{subequations}\label{eq:stationarity}
  \begin{align}\label{eq:stationarity-q}
    q_{0} &= \left\langle
      g_{\mu}^{\prime}(X,h_\eta)^2\right\rangle_\eta\ ,\\  \label{eq:stationarity-chi}
   (\mu_{0} - \beta)\chi_{0} + \beta & = \left\langle \mathbb
        E(h_\eta;X,0)\circ g_{\mu}^{\prime\prime}(h_\eta) \right\rangle_\eta \nonumber \\ & \quad+ \beta \left\langle \mathbb E(h_\eta;X,0)\circ \left[ g_{\mu}^\prime(h_\eta)^2 \right]\right\rangle_\eta \ ,  \end{align}
%\begin{widetext}
 \begin{multline}
  \label{eq:stationarity-mu}
     \frac {\mu_{0}} 4 \chi_{0}\left[ 2(q_{0} + X) + \chi_{0}\right]  =  \frac 1\beta\left\langle \mathbb
        E(h_\eta;X,0)\right. \\ \left. \circ \left\langle \rho_{\mu}(h_{\eta}, \lambda\sqrt{\chi_{0}}) \ln \cosh\left[\beta (h_{\eta} + \lambda\sqrt{\chi_{0}})\right]\right\rangle_{\lambda}\right\rangle_{\eta} \\ - \frac 1\mu_{0} \left\langle \mathbb E(h_\eta;X,0)\circ \ln \left\langle \cosh^{\mu_{0}/\beta}\left[\beta (h_{\eta} + \lambda\sqrt{\chi_{0}})\right]\right\rangle_{\lambda}\right\rangle_{\eta}\ ,
\end{multline}
%\end{widetext}
\end{subequations}
where we denoted
\begin{equation}\label{eq:rho-def}
\rho_{\mu}(h,\lambda\sqrt{\chi_{0}}) = \frac{\cosh\left[\beta (h + \lambda\sqrt{\chi_{0}})\right]^{\mu_{0}/\beta}}{\left\langle \cosh\left[\beta (h + \lambda\sqrt{\chi_{0}})\right]^{\mu_{0}/\beta}\right\rangle_{\lambda}}\ .
\end{equation}
We already mentioned that the individual overlap susceptibilities $\Delta\chi_{l}$ are no longer variational parameters in the free energy with continuous RSB. But their sum $X= K^{-1}\sum_{l} \Delta\chi_{l}$ is. A stationarity equation for this parameter reads 
  \begin{align}\label{eq:stationarity-X}
 X &= \left\langle \mathbb
        E(h_\eta;X,0)\circ \left[ g_{\mu}^\prime(h_\eta)^2 \right]\right\rangle_\eta
    - \left\langle
        g_{\mu}^{\prime}(X,h_\eta)^2\right\rangle_\eta\ .
        \end{align}

Finally, we have to use the stationarity condition for the free energy from Eq.~\eqref{eq:FE-continuous} with respect to infinitesimal variations of the order-parameter function $\mu(x)$. It is easy to find that it can be expressed as                
\begin{align}    \label{eq:stationarity-x}
  x &=   \left\langle \mathbb E(h_\eta;X,0)\circ
        \left[g_{\mu}^\prime(h_\eta)^2\right]\right\rangle_\eta \nonumber \\
    &\quad - \left\langle \mathbb E(h_\eta;X,x)\circ
        \left[g_{\mu}^{\prime}(x,h_\eta)^2\right]\right\rangle_\eta
  \end{align}
valid for any $x \in [0,X]$. Note that
Eq.~\eqref{eq:stationarity-x} for $x =0$ is trivial and for
$x=X$ coincides with Eq.~\eqref{eq:stationarity-X}. Hence, only
equations for $0<x<X$ bring new information for the determination of $\mu(x)$.

There is no explicit dependence on $\mu(x)$ in Eq.~\eqref{eq:stationarity-x}. It is hence difficult to use it for the determination of this order-parameter function. Since this equation holds for any $x$, also its derivative with respect to $x$ must equally hold. The derivative reads
\begin{equation}\label{eq:marginal_{stability}}
1 =  \left\langle \mathbb E(h_\eta;X,x)\circ\left[ g_{\mu}^{\prime\prime}(x,h_{\eta})^{2}\right]\right\rangle_{\eta}\ ,
\end{equation}
which is just a condition for marginal stability of the solution with continuous RSB.\cite{Janis08b}  We recall that the prime stands for the derivative with respect to the magnetic field.  

Neither here we see function $\mu(x)$ explicitly. Yet another differentiation is needed to obtain an explicit expression for the order-parameter function being
\begin{equation}\label{eq:mu_{x}}
\mu(x) = \frac{\left\langle \mathbb E(h_\eta;X,x)\circ\left[ g_{\mu}^{\prime\prime\prime}(x,h_{\eta})^{2}\right]\right\rangle_{\eta}}{2 \left\langle \mathbb E(h_\eta;X,x)\circ\left[ g_{\mu}^{\prime\prime}(x,h_{\eta})^{3}\right]\right\rangle_{\eta}} \ .
\end{equation}
This equation will prove suitable for approximate evaluations. It is also possible to use another derivative and to obtain an explicit equation for $\dot{\mu}(x) = d \mu(x)/dx$ that corresponds to $-dx_{P}/dq_{P}$ in the Parisi notation and is interpreted as the probability distribution of pure states.\cite{Parisi83}  We obtain for this derivative
\begin{widetext}
\begin{multline}\label{eq:dotmu-x}
2\dot{\mu}(x)\left\langle \mathbb E(h_\eta;X,x)\circ\left[ g_{\mu}^{\prime\prime}(x,h_{\eta})^{3}\right]\right\rangle_{\eta}  = - \left\langle \mathbb E(h_\eta;X,x)\circ\left[ g_{\mu}^{(iv)}(x,h_{\eta})^{2}\right]\right\rangle_{\eta} \\
+ 12 \mu(x)  \left\langle \mathbb E(h_\eta;X,x)\circ\left[ g_{\mu}^{\prime\prime}(x,h_{\eta}) g_{\mu}^{\prime\prime\prime}(x,h_{\eta})^{2}\right]\right\rangle_{\eta} 
- 6 \mu(x)^{2}\left\langle \mathbb E(h_\eta;X,x)\circ\left[ g_{\mu}^{\prime\prime}(x,h_{\eta})^{4}\right]\right\rangle_{\eta} \ .
\end{multline}
\end{widetext}
We see that the second term on the right-hand side has the opposite sign from the others and we hence cannot generally guarantee negativity of $\dot{\mu}(x)$. 

To complete the stationarity equations we add explicit expressions for the derivatives of the initial interacting free energy $g_{\mu}(h)$.  With the above introduced notation we have
\begin{subequations}\label{eq:g-derivatives}
\begin{align}\label{eq:g-prime}
g_{\mu}^{\prime} (h) &= \left\langle \rho_{\mu}(h,\lambda\sqrt{\chi_{0}}) t(h_{\lambda})\right\rangle_{\lambda}, 
\end{align}
\begin{widetext}
\begin{align}
\label{eq:g-twoprimes}
g_{\mu}^{\prime\prime} (h)& = \beta \left\langle \rho_{\mu}(h,\lambda\sqrt{\chi_{0}}) \left( 1 - t(h_{\lambda})^{2}\right)\right\rangle_{\lambda}  + \mu_{0} \left[\left\langle \rho_{\mu}(h,\lambda\sqrt{\chi_{0}}) t(h_{\lambda})^{2}\right\rangle_{\lambda}
-  \left\langle \rho_{\mu}(h,\lambda\sqrt{\chi_{0}}) t(h_{\lambda})\right\rangle_{\lambda}^{2}\right]\ ,
\end{align}
\begin{multline}\label{eq:g-threeprimes}
g_{\mu}^{\prime\prime\prime} (h) = 2 \left\langle \rho_{\mu}(h,\lambda\sqrt{\chi_{0}}) t(h_{\lambda})\right\rangle_{\lambda}\left[ \mu_{0}^{2} \left\langle \rho_{\mu}(h,\lambda\sqrt{\chi_{0}}) t(h_{\lambda})\right\rangle_{\lambda}^{2} - \beta^{2} \right] + \left(2\beta^{2} - 3 \beta\mu_{0} + \mu_{0}^{2}\right) \left\langle \rho_{\mu}(h,\lambda\sqrt{\chi_{0}}) t(h_{\lambda})^{3}\right\rangle_{\lambda}
\\  + 3 \mu_{0}(\beta - \mu_{0})  \left\langle \rho_{\mu}(h,\lambda\sqrt{\chi_{0}}) t(h_{\lambda})\right\rangle_{\lambda}  \left\langle \rho_{\mu}(h,\lambda\sqrt{\chi_{0}}) t(h_{\lambda})^{2}\right\rangle_{\lambda}\ .
\end{multline}
\end{widetext}
\end{subequations}
We denoted  $t(h_{\lambda}) \equiv \tanh[\beta(h + \lambda\sqrt{\chi_{0}})]$. Combining the above equations and the definitions for the derivatives of free energy $g_{\mu}$ we can transform the equation for the overlap susceptibility $\chi_{0}$ to
\begin{multline}\label{eq:chi-equation}
\chi_{0} = \left\langle \mathbb E(h_\eta;X,0)\circ\left[ \left\langle \rho_{\mu}(h_{\eta},\lambda\sqrt{\chi_{0}}) t(h_{\lambda,\eta})^{2}\right\rangle_{\lambda}\right]\right\rangle_{\eta} \\
- \left\langle \mathbb E(h_\eta;X,0)\circ\left[ \left\langle \rho_{\mu}(h_{\eta},\lambda\sqrt{\chi_{0}}) t(h_{\lambda,\eta})\right\rangle_{\lambda}^{2}\right]\right\rangle_{\eta}\ ,
\end{multline}
where we abbreviated $h_{\eta,\lambda} = h + \eta\sqrt{q_{0}} + \lambda\sqrt{\chi_{0}}$.

The stationarity equations fully determine equilibrium states of the mean-field free energy with a continuous order-parameter function. These equations are not solvable in their full exact form, being a consequence of inability to solve the Parisi nonlinear partial differential equation~\eqref{eq:g0-DE}. Before we resort to approximations we can use the above representation of the free energy and derive exact representations for the equilibrium values of interesting physical quantities. We can do that in the standard way of statistical mechanics without referring to replicas and the discrete representations used in the derivation of the final form of the free energy with a continuous order-parameter function.

We first evaluate the homogeneous magnetic susceptibility. If we use the condition for marginal stability, Eq.~\eqref{eq:marginal_{stability}}, we obtain        
\begin{align}\label{eq:susceptibility}
  \chi_T &= \left \langle  g''(X,h_{\eta})
  \right\rangle_\eta \nonumber \\
  & = \beta(1 - q_{0}  - \chi_{0} - X)  + \mu_{0} \chi_{0} + \int_{0}^{X}dx \mu(x)\ .
\end{align}
It was argued that $\chi_{T} = 1$ in the glassy phase.\cite{Sompolinsky81} This cannot be deduced from this exact representation without further reasoning.   

As a next interesting thermodynamic quantity we evaluate entropy. For this purpose we have to calculate the temperature derivative of the initial free energy $g_{\mu}$. With the above notation we easily derive
\begin{multline}\label{eq:dgmuT}
\frac{\partial g_{\mu}(h_{\eta})}{\partial T} =\left\langle \rho_{\mu}(h_{\eta,}, \lambda\sqrt{\chi_{0}}) \ln \cosh\left[ \beta(h_{\eta,\lambda}) \right]\right\rangle_{\lambda} \\ - \beta\chi_{0} \left[\beta + (\mu_{0} - \beta) \left\langle \rho_{\mu}(h_{\eta}, \lambda\sqrt{\chi_{0}}) t(h_{\eta,\lambda})^{2}\right\rangle_{\lambda}\right] \\ - \beta h_{\eta} \left\langle \rho_{\mu}(h_{\eta}, \lambda\sqrt{\chi_{0}}) t(h_{\eta,\lambda})\right\rangle_{\lambda}\ .
\end{multline}

Using again the condition for marginal stability and the equations for the equilibrium values of the scalar order parameters we end up with an expression for entropy
\begin{multline}\label{eq:Entropy}
S(h,T) = -\ \frac{\partial f(h,T)}{\partial T}  =  \ln 2 + \beta \chi_{0}(\beta - \mu_{0})(X + q_{0} + \chi_{0})  \\ + \left\langle \mathbb E(h_{\eta};X,0)\circ \left\langle \rho_{\mu}(h_{\eta}, \lambda\sqrt{\chi_{0}})  \ln \cosh\left[\beta (h_{\eta,\lambda})\right]\right\rangle_{\lambda}\right\rangle_{\eta}\\ - \beta \left\langle \mathbb E(h_{\eta};X,0)\circ \left[h \left\langle \rho_{\mu}(h_{\eta}, \lambda\sqrt{\chi}) t(h_{\eta,\lambda})\right\rangle_{\lambda}\right]\right\rangle_{\eta}\\ - \beta^{2}\chi_{0}  - \beta q_{0} \chi_{T} - \frac {\beta^{2}}4 (1 - q_{0} - \chi_{0} - X)^{2} \ .
\end{multline}
The last four terms on the right-hand side of Eq.~\eqref{eq:Entropy} have negative sign and are potentially dangerous for turning the entropy negative at low temperatures.

\section{Perturbation expansion}
\label{sec:PE}

Since we cannot solve the stationarity equations for the order parameters of the free energy  with  continuous RSB, we resort to approximations. We decomposed the full free energy into the zeroth order, corresponding to the 1RSB state, and a correction depending on the continuous order-parameter function $\mu(x)$.  It is natural to formulate a perturbation expansion around the 1RSB solution in powers of function $\mu(x)$. We can do it either directly by using functional derivates and functionals  of $\mu(x)$ or by introducing an interpolation parameter $\xi \in (0,1)$ with which we rescale function $\mu(x)$ and expand all quantities in $\xi$. We choose the first way.  
 
The fundamental quantity to be expanded is the evolution operator
\begin{multline}\label{eq:E-def}
\mathbb E(h_{\eta};X,0)  = \mathbb T_x \exp\left\{ \int_0^X d x \right. \\ \left. 
    \left[\frac 12 \partial_{\bar{h}}^2   + \mu(x) g_{\mu}^{\prime}(x,h_{\eta} +
      \bar{h})\partial_{\bar{h}} \right] \right\} \Bigg|_{\bar{h}=0} \ .
\end{multline}
We have an explicit dependence of this operator on function $\mu(x)$ and an indirect one via the scalar order parameters $q_{0},\chi_{0}, \mu_{0},X$  and function $ g_{\mu}^{\prime}(x,h_{\eta})$. To determine the complete dependence of the evolution operator on $\mu(x)$ we must first evaluate the derivatives with respect to these parameters.  

Dependence on the upper bound $X$ is specific and leads to 
\begin{equation}\label{eq:dE-da}
\frac{\partial \mathbb E(h;X,0)}{\partial X} = \left[\frac 12 \partial_{\bar{h}}^{2} + \mu(X) g_{\mu}^{\prime}(X,h)\partial_{\bar{h}} \right] \mathbb E(h;X,0)\ .
\end{equation}
Partial derivatives with respect to all other scalar parameters have the same generic representation
\begin{subequations}
\begin{multline}\label{eq:dE-dp}
\frac{\partial \mathbb E(h;a,b)}{\partial p} = \int_{b}^{a}d x \mu(x)\\  \mathbb E(h;a,x)
\frac{\partial g_{\mu}^{\prime}(x,h + \bar{h})}{\partial p}\ \partial_{\bar{h}} \mathbb E(h;x,b) \ .
\end{multline}
 From the defining equation for function $ g_{\mu}^{\prime}$, Eq.~\eqref{eq:E-equation}, we obtain
\begin{multline}\label{eq:dgprime-dp}
\frac{\partial g_{\mu}^{\prime}(x,h)}{\partial p} =  \mathbb E(h;x,0)\circ\left[\frac{\partial g_{\mu}^{\prime}(h)}{\partial p}\right]\\ + \int_{0}^{x}d y \mu(y) \mathbb E(h;x,y) \circ \left[g_{\mu}^{\prime\prime}(y,h)
\frac{\partial g_{\mu}^{\prime}(y,h)}{\partial p}\right] \ . 
\end{multline}\end{subequations}

We do the same with the functional derivative with respect to $\mu(x)$, 
\begin{subequations}
\begin{multline}\label{eq:deltaE-deltamu}
\frac{\delta \mathbb E(h;a,b)}{\delta \mu(x)} =  \mathbb E(h;a,x) \circ\left[g_{\mu}^{\prime}(x,h)\partial_{h}\mathbb E(h;x,b)\right] \\
+ \int_{x}^{a}d y \mu(y) \mathbb E(h;a,y) \circ\left[\frac{\delta g_{\mu}^{\prime}(y,h)}{\delta \mu(x)}\partial_{h} \mathbb E(h;y,b)\right]
\end{multline}
and
\begin{multline}\label{eq:deltagprime-deltamu}
\frac{\delta g_{\mu}^{\prime}(x,h)}{\delta \mu(y)} =  \mathbb E(h;x,y)\circ\left[ g_{\mu}^{\prime}(y,h)  g_{\mu}^{\prime\prime}(y,h)\right]\\ + \int_{y}^{x}d u \mu(u) \mathbb E(h;x,u) \circ \left[ \frac{\delta g_{\mu}^{\prime}(u,h)}{\delta \mu(y)}g_{\mu}^{\prime\prime}(y,h)\right] \ .
\end{multline}\end{subequations}

Putting all the derivatives together we can set a basic equation for an iterative determination of the next approximation to the full evolution operator. Knowing the $n$th and $(n-1)$th orders of this operator ${\mathbb E}^{(n)}(h;a,b)$ and ${\mathbb E}^{(n-1)}(h;a,b)$ the next approximation then is 
\begin{widetext}
\begin{multline}\label{eq:E-PE}
{\mathbb E}^{(n+1)}(h;a,b) = {\mathbb E}^{(n)}(h;a,b) \\ + \int_{b}^{a}d x \mu(x)\left[ {\mathbb E}^{(n)}(h;a,x) g_{\mu}^{\prime}(x,h + \bar{h}) \partial_{\bar{h}} {\mathbb E}^{(n)}(h;x,b)  + \delta_{a,X} \frac{\delta X}{\delta \mu(x)} 
\left( \frac 12 \partial_{\bar{h}}^{2} +  \mu(X)g_{\mu}^{\prime}(X,h) \partial_{\bar{h}} \right) {\mathbb E}^{(n)}(h;X,b) \right.  \\ \left.
 +  \int_{0}^{X}d y\mu(y)  {\mathbb E}^{(n-1)}(h;a,x)\left(\sum_{i}\frac{\delta p_{i}}{\delta \mu(y)} \frac{ \partial g_{\mu}^{\prime}(x,h + \bar{h})}{\partial p_{i}}  + \theta(x - y)\frac{\delta g_{\mu}^{\prime}(x,h + \bar{h})}{\delta \mu(y)}\right) \partial_{\bar{h}} {\mathbb E}^{(n-1)}(h;x,b)\right]^{(n)} \ ,
\end{multline}
\end{widetext}
where $p_{i} \in \{q_{0}, \chi_{0}, \mu_{0}\}$ and $\left[F[\mu(x)]\right]^{(n)}$ means that only the $n$th power of function $\mu(x)$ from functional  $F[\mu(x)]$  is taken into account.  That is
\begin{widetext}
\begin{equation*}
\left[F[\mu(x)]\right]^{(n)} = \int dx_{1} \ldots dx_{n}  \mu(x_{1}) \ldots \mu(x_{n}) %\\  \times
\frac{\delta^{n}F[\mu(x)]}{\delta \mu(x_{1}) \ldots \delta \mu(x_{n})}\Bigg|_{\mu(x) = 0} \ .
\end{equation*}
\end{widetext}
The initial evolution operator is 
\begin{align}\label{eq:E0}
{\mathbb E}^{(0)}(h;a,b) & = \exp\left\{\frac 12 (a - b) \partial_{h}^{2} \right\}
\end{align}
and we set ${\mathbb E}^{(-1)} = 0$ to comply with Eq. ~\eqref{eq:E-PE} for $n=0$.  In this way a formal power expansion in the continuous order-parameter function is exhaustively defined.

\section{Lowest-order approximation}
\label{sec:FOA}

 The zeroth-order approximation is the 1RSB free energy with the evolution operator approximated by formula~\eqref{eq:E0}. This evolution operator can be transformed to a Gaussian integral. To make the expressions for the equations within this approximation as compact as possible we introduce the following  generic notation
\begin{subequations}
\begin{align}\label{eq:E-functional}
 E_{x}[f(h)] & = \int_{-\infty}^{\infty} \frac{d \phi}{\sqrt{2\pi}} e^{-\phi^2/2} f(\beta(h + \phi\sqrt{x})) \ , \\
 \left\langle\rho_{\mu}(h) f(h)\right\rangle_{\chi_{0}} & = \frac{E_{\chi_{0}}\left[\cosh (h)^{\mu_{0}/\beta }  f(h)\right] }{E_{\chi_{0}}\left[\cosh (h)^{\mu_{0}/\beta } \right]}\ .
\end{align}
\end{subequations}
 The zeroth-order approximation leads to equations where it is convenient to replace the parameter $q_{0}$ with a new variable  $Y = X + q_{0}$. Replacing the exact evolution operator $\mathbb E$ by the approximate one from Eq.~\eqref{eq:E0} in Eqs.~\eqref{eq:stationarity}
we obtain  
\begin{subequations}\label{eq:1RSB}
\begin{align}\label{eq:Y-equation}
Y & = E_{Y}\left[\left\langle\rho_{\mu}(h) t(h)\right\rangle_{\chi_{0}}^2 \right]\ ,  \\ \label{eq:chi0-equation}
\chi_{0} & =  E_{Y}\left[\left\langle\rho_{\mu}(h) t(h)^2\right\rangle_{\chi_{0}} \right]  - E_{Y}\left[\left\langle\rho_{\mu}(h) t(h)\right\rangle_{\chi_{0}}^2 \right] \ .
\end{align}
Replication index $\mu_{0}$ is determined from
\begin{multline}\label{eq:mu-equation}
 \frac{1}4 \beta\chi_{0} \mu_{0} (2Y + \chi_{0})  = E_Y\left[ \left\langle\rho_\mu(h)\ln \cosh (h) \right\rangle_{\chi_{0}} \right] \\
 - \frac {\beta}{\mu_{0}} E_Y\left[\ln E_{\chi_{0}}\left[\cosh(h)^{\mu_{0}/\beta}\right]\right]\ .
\end{multline}\end{subequations}
The equation for parameter $X$ follows from Eqs.~\eqref{eq:stationarity-q} and~\eqref{eq:1RSB} 
\begin{multline}\label{eq:X-equation}
X =  E_{Y}\left[\left\langle\rho_{\mu}(h) t(h)\right\rangle_{\chi_{0}}^2 \right] \\ -
 E_{(Y-X)}\left[E_{X}\left[\left\langle\rho_{\mu}(h) t(h)\right\rangle_{\chi_{0}}\right]^2 \right] \ .
\end{multline}

Finally, the continuous order-parameter function from Eq.~\eqref{eq:mu_{x}} in the lowest-order approximation is 
\begin{equation}\label{eq:mux-equation}
\mu(x) = \frac{E_{(Y-x)}\left[E_x\left[ g_{\mu}^{\prime\prime\prime}(h) \right]^2 \right]}
{2 E_{(Y-x)}\left[E_x\left[ g_{\mu}^{\prime\prime}(h) \right]^3 \right]}\ .
\end{equation} 

Defining equations~\eqref{eq:1RSB} determine order parameters $\chi_{0}, \mu_{0}, Y$ of a 1RSB state. It can be seen from a generating free energy 
\begin{multline}\label{eq:FE-ZO}
 f_0(\chi_{0}, \mu_{0}, Y) = - \frac  \beta 4 (1 - \chi_{0} - Y)^2  + \frac 14  \mu_{0}\chi_{0}  \left(2Y  + \chi_{0}\right) \\ - \frac 1\beta \ln 2 - \frac 1{\mu_{0}} E_Y\left[\ln E_{\chi_{0}}\left[ \cosh(h)^{\mu_{0}/\beta}\right] \right]
\end{multline}
to which these equations define stationarity points. Free energy $f_{0}$ is just the free energy  from Eq.~\eqref{eq:FE-continuous} with the evolution operator from Eq.~\eqref{eq:E0}.  

Parameter $X$ that does not appear in the free energy, Eq.~\eqref{eq:FE-ZO}, is the first iteration for the length of the interval on which the continuous order-parameter function lives. If $X>0$, then a solution with continuous RSB exists. Note that Eq.~\eqref{eq:X-equation} has two solutions $X = 0$ and $X=Y$ for $h=0$.  We can always take the latter one as the starting point for the perturbation expansion in which the Sherrington-Kirkpatrick (SK) solution is then completely circumvented ($q_{SK}=0$). It means that we always can construct a solution with continuous replica-symmetry breaking for $h=0$ independently of whether  an equilibrium state with finite-many replica hierarchies is locally stable or not.  In an applied magnetic field, a non-zero parameter $X$ generally exists if an instability condition is satisfied
\begin{align}\label{eq:X-instability}
E_{Y}\left[ g_{\mu}^{\prime\prime}(h)^{2}\right] & > 1
\end{align}
with $g_{\mu}^{\prime\prime}(h)$ from Eq.~\eqref{eq:g-twoprimes} and parameters $Y$, $\chi_{0}$ and $\mu_{0}$ being solutions of Eqs.~\eqref{eq:1RSB}.  It is sufficient if condition~\eqref{eq:X-instability} is satisfied for any of the solutions of Eqs.~\eqref{eq:1RSB}, that is, either the paramagnetic, replica-symmetric or one-level replica-symmetry-breaking solution.      

Parameters $Y,\chi_{0},\mu_{0}$ determined from Eqs.~\eqref{eq:1RSB} define a stationarity point of the free energy from Eq.~\eqref{eq:FE-ZO} with one level of replica symmetry breaking. Parameter $X$ from Eq.~\eqref{eq:X-equation} and function $\mu(x)$ from Eq.~\eqref{eq:mux-equation} go beyond the 1RSB solution and represent the leading-order correction to the 1RSB approximation towards a solution with continuous RSB. Parameter $X$ determines an interval $[0,X]$ on which the continuous order-parameter function $\mu(x)$ is defined.  Both quantities are determined within the 1RSB approximation.      
 
We must go to the next iteration ${\mathbb E}^{(1)}$ of the evolution operator to obtain corrections to the 1RSB results. We find from Eq.~\eqref{eq:E-PE}
\begin{multline}\label{eq:E1}
\left\langle {\mathbb E}^{(1)}(h_{\eta};X,0)\circ f(h_{\eta})\right\rangle_{\eta} = E_{Y}[f(h)]\\  + \int_{0}^{X}d x \mu(x) \left\{\frac 12 \frac{\delta X}{\delta\mu(x)} E_{Y}[f^{\prime\prime}(h)] \right. \\ \left.    
\phantom{\frac 12}+ E_{(Y-x)}\left[ E_{x}\left[g_{\mu}^{\prime}(h)\right][ E_{x}\left[f^{\prime}(h)\right]\right]\right\} \ .
\end{multline}
To make the representation in Eq.~\eqref{eq:E1} explicit we need to determine the functional derivative $\delta X/\delta\mu(x)$.  We add  a first-order correction to the 1RSB free energy, which is
\begin{multline}\label{eq:Delta-FE}
\Delta f_1 = \frac 12 \int_{0}^{X}dx \mu(x) \bigg\{Y - x  \\
 -  E_{(Y-x)}\left[E_{x}\left[ g^{\prime}_{\mu}(h)\right]^2 \right] \Bigg\}\ .
\end{multline}
If we now use free energy $f_{1} = f_{0} + \Delta f_{1}$ as a generating functional for parameters $\chi_{0}, \mu_{0}, Y, X$  with $\mu(x)$ as an external source, the stationarity equations of this free energy define the scalar order parameters with their first correction due to the continuous order-parameter function $\mu(x)$ obtained from Eq.~\eqref{eq:mux-equation} evaluated within the 1RSB approximation.  

If we want to evaluate the first correction to function $\mu(x)$ from Eq.~\eqref{eq:mux-equation} we have to go further and to calculate the free energy to the second order in $\mu(x)$. The second correction to the 1RSB free energy reads
\begin{multline}\label{eq:Delta2-FE}
\Delta f_2  = - \int_{0}^{X} dx \int_{0}^{x} dy \mu(x)\mu(y)  
 E_{(Y-x)}\left[E_{x}\left[ g^{\prime}_{\mu}(h)\right]\right. \\ \left. E_{(x-y)}\left[E_{y}\left[g^{\prime}_{\mu}(h) \right] E_{y}\left[g^{\prime\prime}_{\mu}(h) \right]\right] \right] \ .
\end{multline}
If we now add both corrections to the 1RSB free energy we obtain a new free energy, exact to second order in $\mu(x)$, that has the following representation 
\begin{widetext}
\begin{multline}\label{eq:f2}
f_{2}(\chi_{0},\mu_{0},X,Y;\mu(x)) = f_{0} + \Delta f_{1} + \Delta f_{2} \\ = - \frac  \beta4 (1 - \chi_{0} - Y)^2  + \frac 14  \mu_{0}\chi_{0}\left(2Y   + \chi_{0}\right) - \frac 1{\mu_{0}} E_Y\left[\ln E_{\chi_{0}}\left[ \left(2 \cosh\left( h\right)\right)^{\mu_{0}/\beta}\right] \right]  + \frac 12 \int_{0}^{X}\!\!\!  dx \mu(x) \bigg\{Y - x \\ -  E_{(Y-x)}\left[E_{x}\left[ g^{\prime}_{\mu}(h)\right]^2 \right] \Bigg\} - \int_{0}^{X} \!\!\!  dx \mu(x) \!\!\int_{0}^{x} \!\! dy \mu(y)  
 E_{(Y-x)}\left[E_{x}\left[ g^{\prime}_{\mu}(h)\right]  E_{(x-y)}\left[E_{y}\left[g^{\prime}_{\mu}(h) \right] E_{y}\left[g^{\prime\prime}_{\mu}(h) \right]\right] \right]\ .   
\end{multline}
To derive the correction to $\mu(x)$ from Eq.~\eqref{eq:mux-equation} we proceed in the same way as in the exact case. Setting zero the first variation of free energy $f_{2}$ with respect to  $\mu(x)$ leads to
\begin{multline}\label{eq:mu-x-0}
\frac 12\left\{ Y - x  -  E_{(Y-x)}\left[E_{x}\left[ g^{\prime}_{\mu}(h)\right]^2 \right] \right\} = \int_{0}^{x} \!\! dy \mu(y)  
 E_{(Y-x)}\left[E_{x}\left[ g^{\prime}_{\mu}(h)\right]  E_{(x-y)}\left[E_{y}\left[g^{\prime}_{\mu}(h) \right] E_{y}\left[g^{\prime\prime}_{\mu}(h) \right]\right] \right]   \\
+ \int_{x}^{X} \!\! dy \mu(y)  
 E_{(Y-y)}\left[E_{y}\left[ g^{\prime}_{\mu}(h)\right]  E_{(y-x)}\left[E_{x}\left[g^{\prime}_{\mu}(h) \right] E_{x}\left[g^{\prime\prime}_{\mu}(h) \right]\right] \right] \ . 
\end{multline}
Its derivative with respect to $x$ results in
\begin{multline}\label{eq:mu-x-1}
1 = E_{(Y-x)}\left[E_{x}\left[g^{\prime\prime}_{\mu}(h)\right]^{2}\right] + 2\int_{0}^{x}dy \mu(y)  E_{(Y-x)}\left[E_{x}\left[ g^{\prime\prime}_{\mu}(h)\right] E_{(x-y)}\left[E_{y}\left[g^{\prime\prime}_{\mu}(h)\right]^{2}  + E_{y}\left[g^{\prime}_{\mu}(h)\right] E_{y}\left[g^{(\prime\prime\prime)}_{\mu}(h)\right]\right]\right] \\
 + 2\int_{x}^{X}dy \mu(y) E_{(Y-y)}\left[E_{y}\left[ g^{\prime}_{\mu}(h)\right] E_{(y-x)}\left[E_{x}\left[ g^{\prime\prime}_{\mu}(h)\right] E_{x}\left[g^{\prime\prime\prime}_{\mu}(h)\right]\right]\right]
\end{multline}
 and the second derivative leads to an equation from which we obtain the desired correction to the order-parameter function
 \begin{multline}\label{eq:mu-x-2}
2\mu(x) E_{(Y-x)}\left[E_{x}\left[g^{\prime\prime}_{\mu}(h)\right]^{3}\right] = E_{(Y-x)}\left[E_{x}\left[g^{\prime\prime\prime}_{\mu}(h)\right]^{2}\right] \\+ 2\int_{0}^{x}dy \mu(y)  E_{(Y-x)}\left[E_{x}\left[ g^{\prime\prime\prime}_{\mu}(h)\right] E_{(x-y)}\left[3E_{y}\left[g^{\prime\prime}_{\mu}(h)\right] E_{y}\left[g^{\prime\prime\prime}_{\mu}(h)\right] %\right. \right. \\ \left. \left.
 + E_{y}\left[g^{\prime}_{\mu}(h)\right] E_{y}\left[g^{(iv)}_{\mu}(h)\right]\right]\right] \\
 + 2\int_{x}^{X}dy \mu(y) E_{(Y-y)}\left[E_{y}\left[ g^{\prime}_{\mu}(h)\right] E_{(y-x)}\left[E_{x}\left[ g^{\prime\prime\prime}_{\mu}(h)\right] E_{x}\left[g^{(iv)}_{\mu}(h)\right]\right]\right]\ .
\end{multline}
\end{widetext}
Knowing the corrections to all the order parameters, free energy  of the 1RSB solution and to the evolution operator we can evaluate  corrections also to other physical quantities, in particular magnetic susceptibility $\chi_{T}$ and entropy $S(T)$ by applying our approximation to the exact formulas, Eq.~\eqref{eq:susceptibility} and Eq.~\eqref{eq:Entropy}.   

Approximating all physical quantities by expanding the evolution operator in the power series from Eq.~\eqref{eq:E-PE} in the exact equations for their equilibrium values makes the approximate theory only approximately thermodynamically consistent. That is, entropy calculated from Eq.~\eqref{eq:Entropy} with an approximate evolution  operator ${\mathbb E}^{(1)}$ does not coincide with the entropy calculated from the temperature derivative of free energy $f_{1} = f_{0} + \Delta f_{1}$, since function $\mu(x)$ is treated in the latter as an external source that depends via Eq.~\eqref{eq:mux-equation} on temperature and, consequently, the two definitions do not coincide. It means that thermodynamic consistency is obeyed by approximate quantities only to one order  lower  than that chosen in the evolution operator. This deficiency can be removed if we had a free energy being stationary also with respect to infinitesimal fluctuations of function $\mu(x)$. Such a free energy is $f_{2}$ from Eq.~\eqref{eq:f2}. It can be treated as a generating functional for all its variables, including $\mu(x)$. The continuous order-parameter function can no longer be treated as a perturbation but is rather determined self-consistently with other scalar order parameters from Eq.~\eqref{eq:mu-x-0}.  Such an approximation would be fully thermodynamically consistent and exact up to the second order in $\mu(x)$ for all physical quantities.  

\section{Results}
\label{sec:Results}
%%%%%%%%%%%%%%%%%%%%%%%%%%%%%%%%%%%%%%%

The starting point of the presented perturbation expansion, the zeroth order approximation, is a solution with one level of replica-symmetry breaking (1RSB), the free energy of which with its order parameters is given in Eq.~\eqref{eq:FE-ZO}. The replica symmetric solution is part of this approximation if we put $\chi_{0} =0$ and neglect the stationarity equation for this parameter. Alternatively, due to degeneracy of the solution,\cite{Janis05c} we can choose $\mu_{0} = \beta$. Then, free energy $f_{0}$ from Eq.~\eqref{eq:FE-ZO} becomes independent of susceptibility $\chi_{0}$, which also leads to the replica-symmetric solution. The two ways to reproduce the replica-symmetric solution correspond to different identifications of the Sherrington-Kirkpatrick parameter $q_{SK}$ with either $\chi_{0}$ or $Y$ in our notation. In the former case we have $\chi_{0}=0$ and $Y = q_{SK}>0$, while in the latter $Y=0, \mu_{0}= 0$, and $\chi_{0}= q_{SK}>0$ for $T< T_{c}=1$. The full 1RSB solution is obtained so that parameters $\chi_{0}$ and $Y$ are calculated self-consistently for each value of $\mu_{0}$. The latter parameter is then determined from the local maximum of free energy $f_{0}$. Without magnetic field we have another degeneracy in the 1RSB equation~\eqref{eq:X-equation}, allowing us to choose $X=Y$. We use the values of  parameters $Y,\chi_{0}$ and $\mu_{0}$ from Eqs.~\eqref{eq:1RSB} to determine the continuous order-parameter function $\mu(x)$ from Eq.~\eqref{eq:mux-equation}.   In this way we completed the starting approximation to which we can evaluate  corrections in powers of $\mu(x)$.

We calculated explicitly only linear corrections in function $\mu(x)$ to the 1RSB results.  In the perturbation expansion around the 1RSB solution for $h=0$  we obtain  $X=Y$ in all orders.  Corrections to the parameters from the 1RSB solution are calculated from stationarity equations for free energy $f_{1} =f_{0} + \Delta f_{1}$ from Eq.~\eqref{eq:Delta-FE} with respect to infinitesimal variations of $Y, \chi_{0}$ and $\mu_{0}$ and with an external source $\mu(x)$ determined from Eq.~\eqref{eq:mux-equation} with its parameters fixed at the 1RSB values.  Or, equivalently, one can use  the evolution operator $\mathbb{E}^{(1)}$ from Eq.~\eqref{eq:E1} in the exact equations for the order parameters from the exact free energy. 

For comparison we also applied the expansion in the continuous order-parameter to the replica-symmetric solution. Due to the degeneracy in the identification of the SK solution, we must choose the one that is more unstable. It appears that the SK solution with $\chi_{0} = q_{SK}$ leads to $Y=X =0$ in all orders and we generate no corrections to the replica-symmetric solution. The other limit to the SK solution with $\chi_{0}=0$ leads to $X =Y > 0$, $q_{0}=0$ in all orders of the perturbation expansion. It means that an expansion around the SK solution coincides with an expansion around the paramagnetic solution with no SK order parameter.          

Simultaneously with the perturbation expansion we used Monte Carlo simulations on fully connected graphs with up to $N=512$ spins with different number of Monte-Carlo sweeps. Usually we used 4096x256 equilibrating steps followed by  around one million Monte-Carlo steps from which we registered  the data for averaging  after each 16-256 steps for each configuration of the exchange couplings out of 6x1024 random selections.\cite{Mackenzie82}       

\begin{figure}
\includegraphics[width=0.48\textwidth]{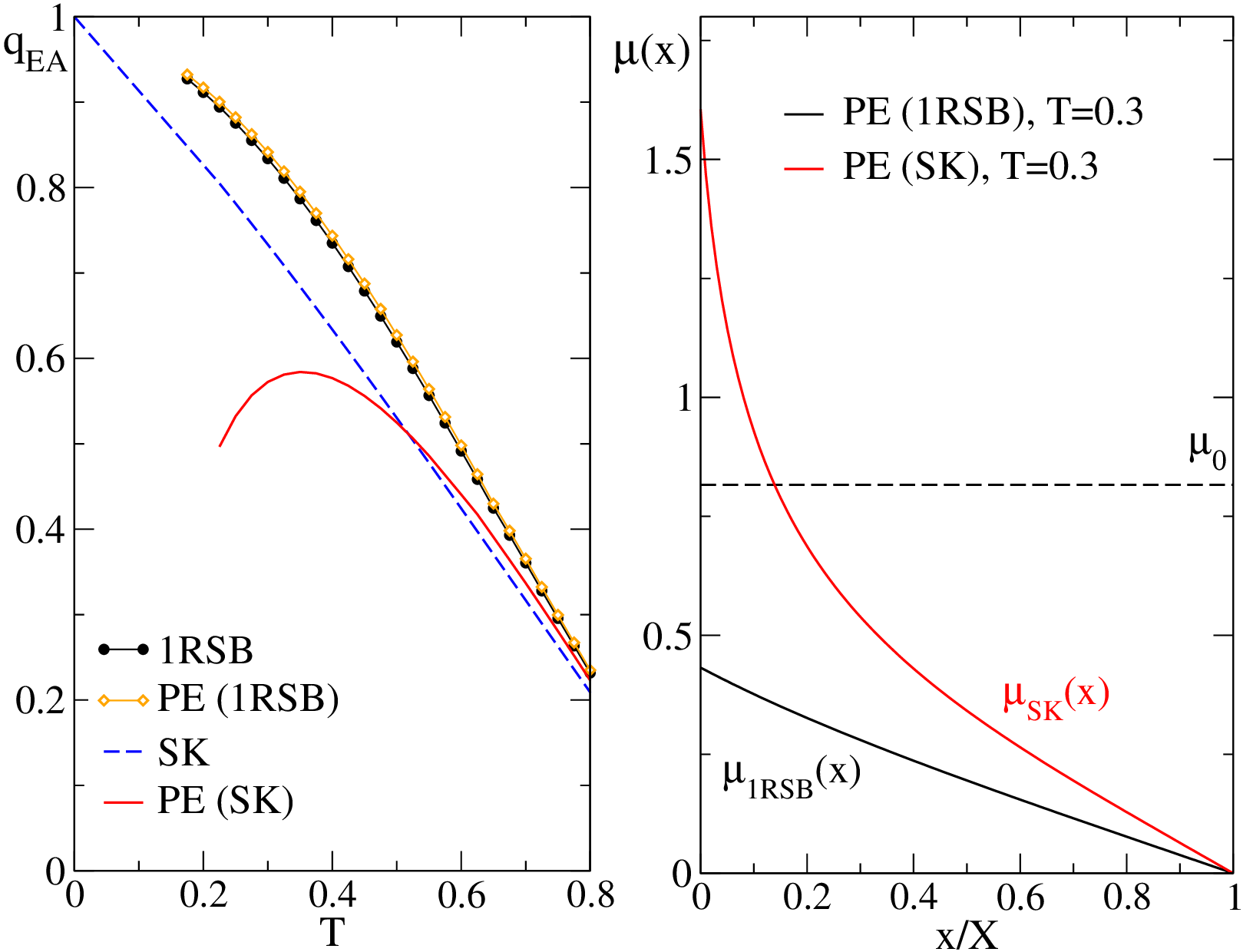}
\caption{(Color online) Edwards-Anderson order parameter $q_{EA}$ (left panel) and the continuous order-parameter function  $\mu(x)$ (right panel) calculated using perturbation expansions (PE) around the 1RSB and replica-symmetric (SK) solutions. Parameter $\mu_0$ of 1RSB is plotted for reference.}
\label{fig:qEA_with_SK}
\end{figure}
\begin{figure}
\includegraphics[width=0.48\textwidth]{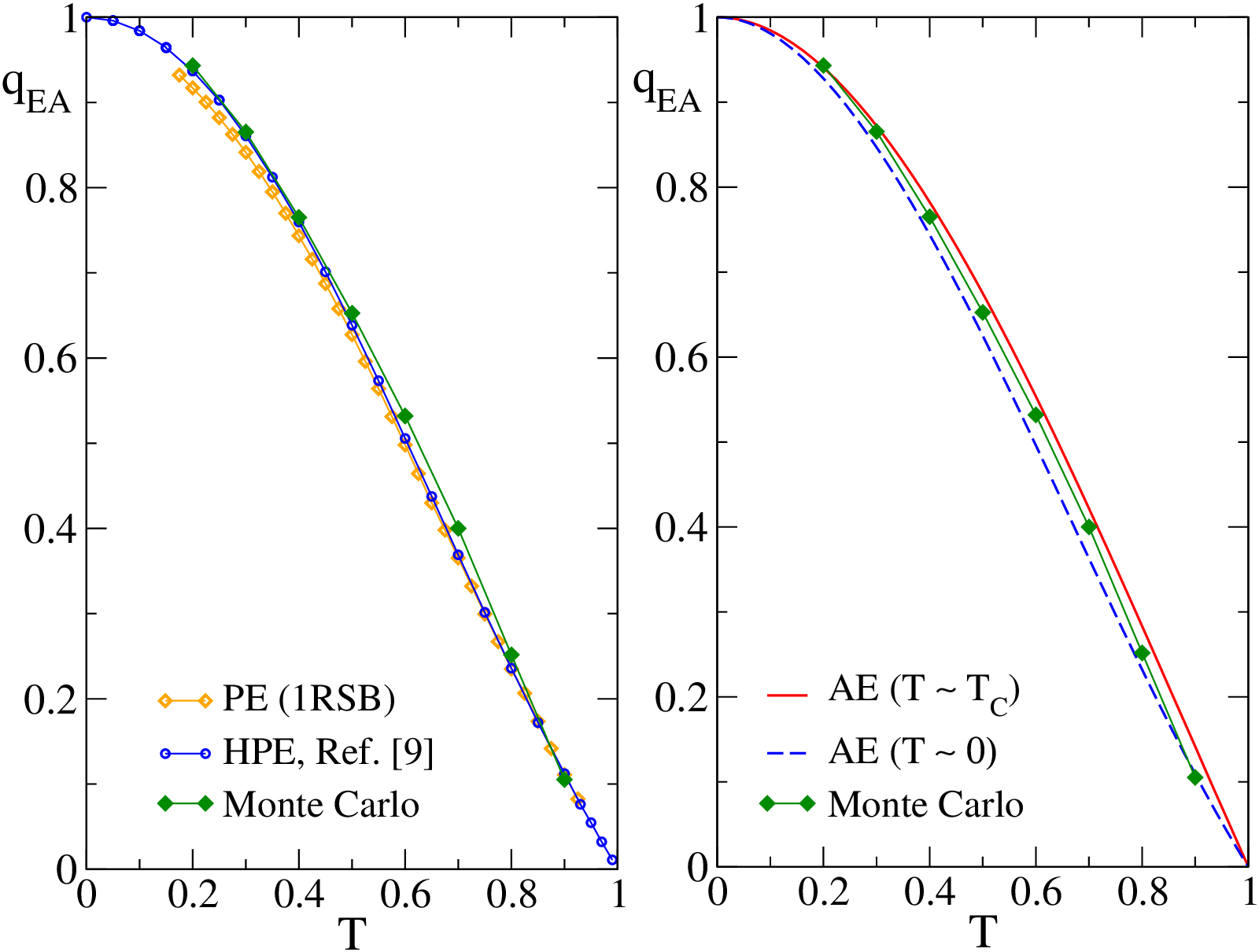}
\caption{(Color online) Temperature dependence of the Edwards-Anderson order parameter resulting from different approximations. Left panel: Edwards-Anderson parameter from the 1RSB solution with first order correction (PE (1RSB)), compared with the solution of Ref.~[\onlinecite{Crisanti02}] (HPE) and Monte Carlo simulations. Right panel: Monte Carlo result compared with extrapolations of asymptotic expansions around the transition temperature $T_{c}=1$  (solid line) and zero temperature (dashed line). }
\label{fig:qEA_all}
\end{figure}
We plotted in Fig.~\ref{fig:qEA_with_SK} the Edwards-Anderson (EA) order parameter, being in our notation $q_{EA}= \chi_{0} + Y$, calculated in the replica-symmetric approximation, 1RSB solution and from the first-order of the perturbation expansion (PE)  around either of these solutions. We can see that the perturbation expansion in the first order gives almost no correction to the 1RSB approximation, while there are tangible changes due to the perturbation expansion when applied to the SK solution. Even more, at lower temperatures ($T\approx 0.4$) the first-order correction to parameter $X$, that equals $q_{SK}$ in the lowest order,   starts to decrease and wrongly downturns the slope of the EA parameter. The reason for this unreliable behavior of the first correction to the SK solution is the value of the ``small parameter'' $\mu(x)$ used. The right panel of  Fig.~\ref{fig:qEA_with_SK} displays $\mu(x)$ at $T=0.3$ calculated in the 1RSB and SK solutions. The function from the SK solution is significantly higher, making thus the perturbation expansion around the SK (paramagnetic) solution much less reliable than the one around the 1RSB solution. It is understandable, since the role of the continuous order-parameter function $\mu(x) $ is partly overtaken by the scalar value $\mu_{0}$ in the 1RSB approximation.         Unlike function $\mu(x)$, parameter $\mu_{0}$ is treated non-perturbatively.

To check reliability of the perturbation expansion we compared the Edwards-Anderson order parameter in Fig.~\ref{fig:qEA_all}, left panel, with the one obtained from Monte-Carlo simulations and the high-order perturbation expansion (HPE)  of Crisanti and Rizzo, Ref.~[\onlinecite{Crisanti02}].  We can see that there is not a big difference between the perturbation expansion around the 1RSB solution and Monte-Carlo simulations and the expansion of Crisanti and Rizzo for temperatures $T> 0.3$.  As one expects, a better precision in  lower temperatures demands inclusion of higher orders of the expansion.  As a curiosity we compared (right panel) the Edwards-Anderson  parameter from Monte-Carlo simulations with low-orders of two asymptotic expansions, one, high-temperature, around the transition temperature and the other, low-temperature, around zero temperature. The former expansion is\cite{Janis06b,Young83}  $q_{EA} \doteq 1 - 2T^{2} + T^{3}$ while the latter to the same order is\cite{Crisanti02} $q_{EA}\doteq 1 - 1.6 T^{2} + 0.6 T^{3}$. We see that the Monte-Carlo data are surprisingly well reproduced by the expansions in the respective regions and do not differ much from them in the whole glassy phase. It is worth noting that the expansions are only asymptotic and higher terms destroy this surprising reconstruction of the Monte-Carlo Edwards-Anderson parameter.          
\begin{figure}
\includegraphics[width=0.48\textwidth]{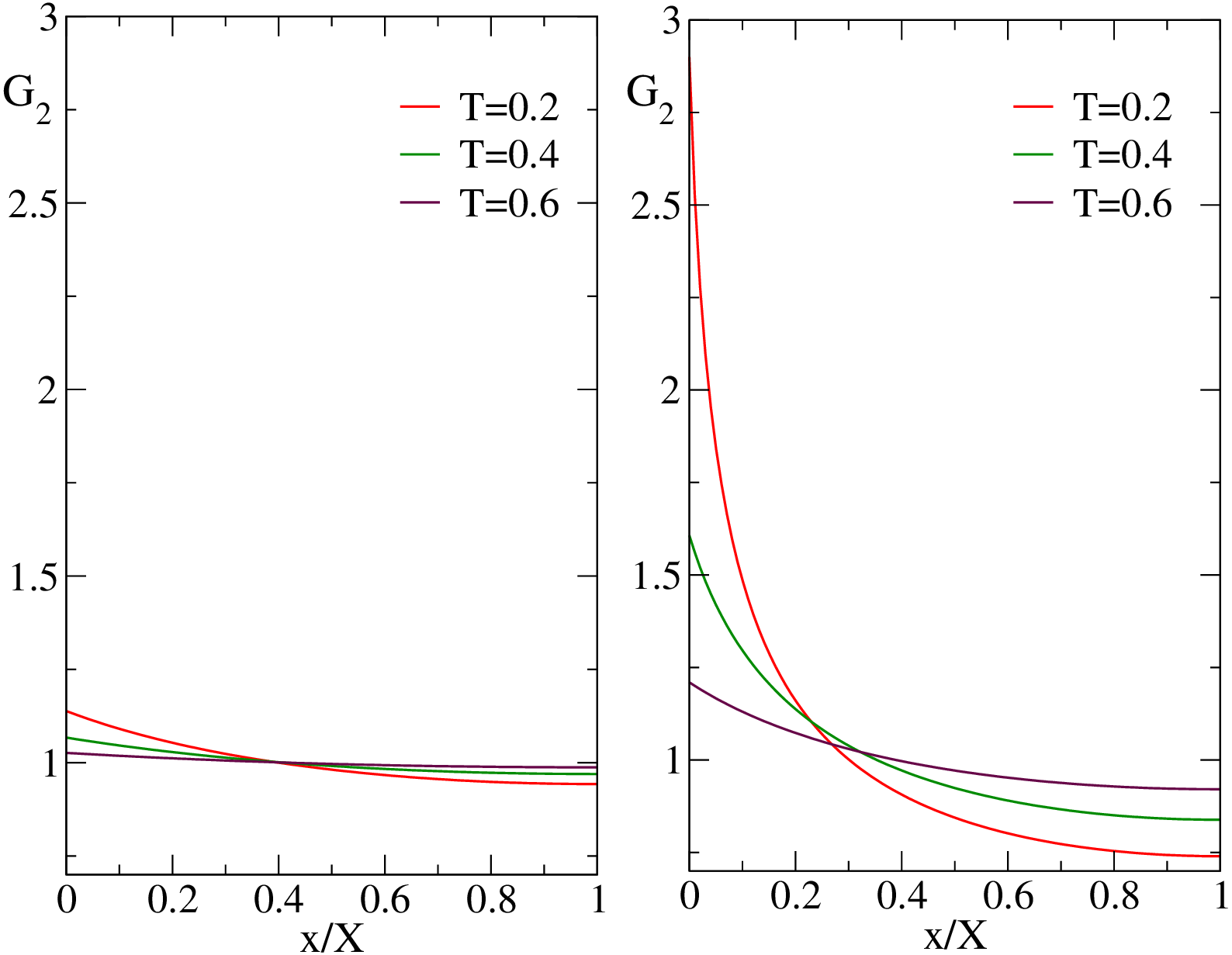}
\caption{(Color online) Stability condition, right-hand side of Eq.~\eqref{eq:marginal_{stability}}, denoted $G_{2}$, for the expansion around 1RSB (left panel) and around SK (right panel).  The curves from top to bottom at $x=0$ correspond to temperatures $T=0.2$, $T=0.4$, $T=0.6$, respectively.}
\label{fig:g2_vs_x}
\end{figure}

An important property of the full Parisi solution is the condition of (marginal) stability, Eq.~\eqref{eq:marginal_{stability}}, that may be used as one of criteria of reliability and consistency of approximations. We plotted in Fig.~\ref{fig:g2_vs_x} the right-hand side of Eq.~\eqref{eq:marginal_{stability}}, denoted by $G_{2}$, for the first-order expansion around the 1RSB solution (left panel) and the SK solution (right panel) and different temperatures.  Both approximate solutions worsen their reliability with lowering the temperature but the one around the SK solution shows much larger deviations from the stability value $G_{2}=1$. The expansion around the 1RSB solution, on the other hand, does not violate the marginal stability significantly for all temperatures $T>0.2$ and gives a good estimate of the behavior of the full solution with continuous replica-symmetry breaking.      

\begin{figure}%\vspace{10pt}
\includegraphics[width=0.48\textwidth]{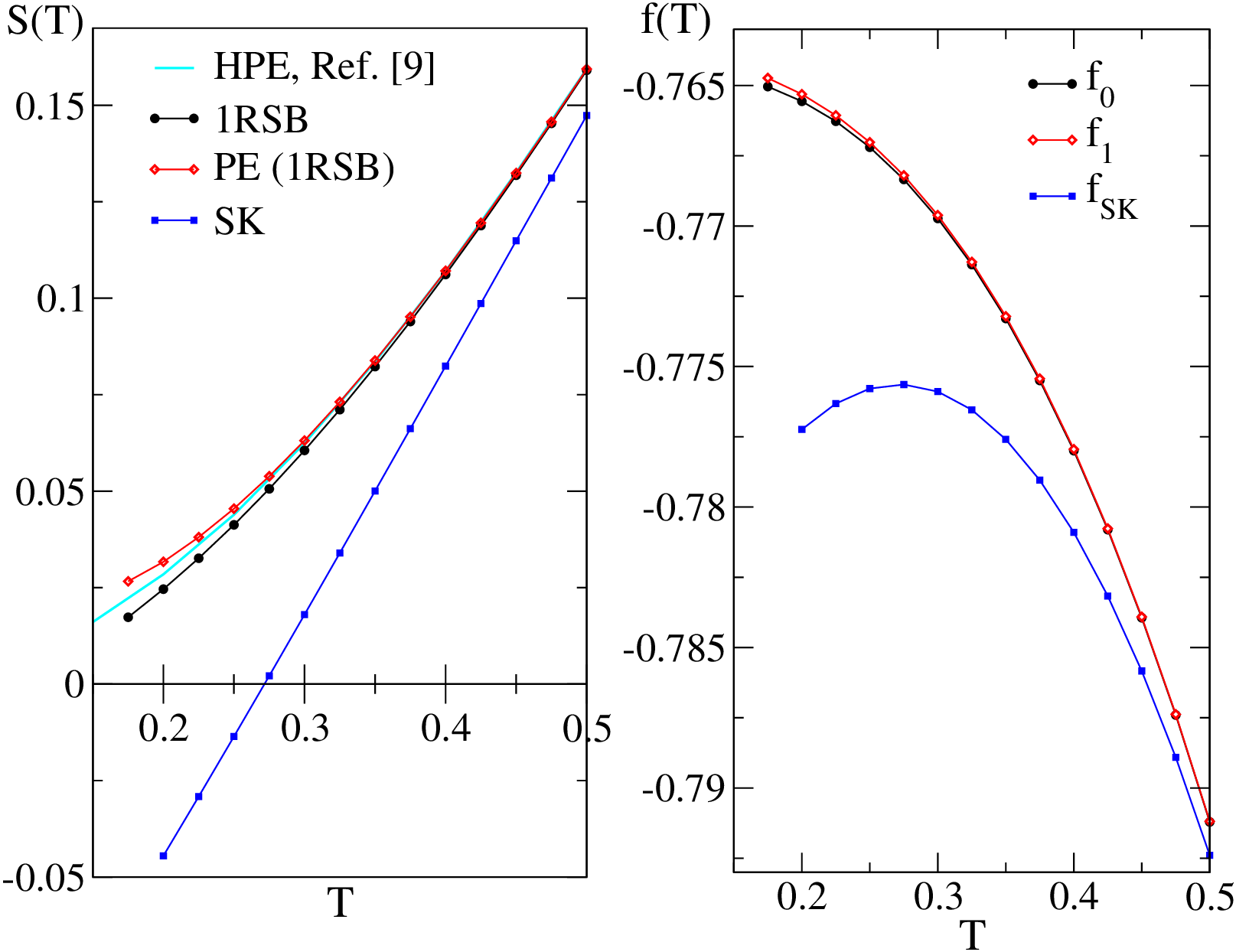}
\caption{(Color online) Temperature dependence of entropy (left panel) calculated in 1RSB, with first correction to it, and the replica-symmetric solution. We also used the data for entropy  from Ref.~[\onlinecite{Crisanti02}]. The right panel shows free energy of the SK solution ($f_{SK}$), 1RSB ($f_{0}$) and the first correction to 1RSB ($f_{1}$). }
\label{fig:entr_ff}
\end{figure}
\begin{figure}
\includegraphics[width=0.48\textwidth]{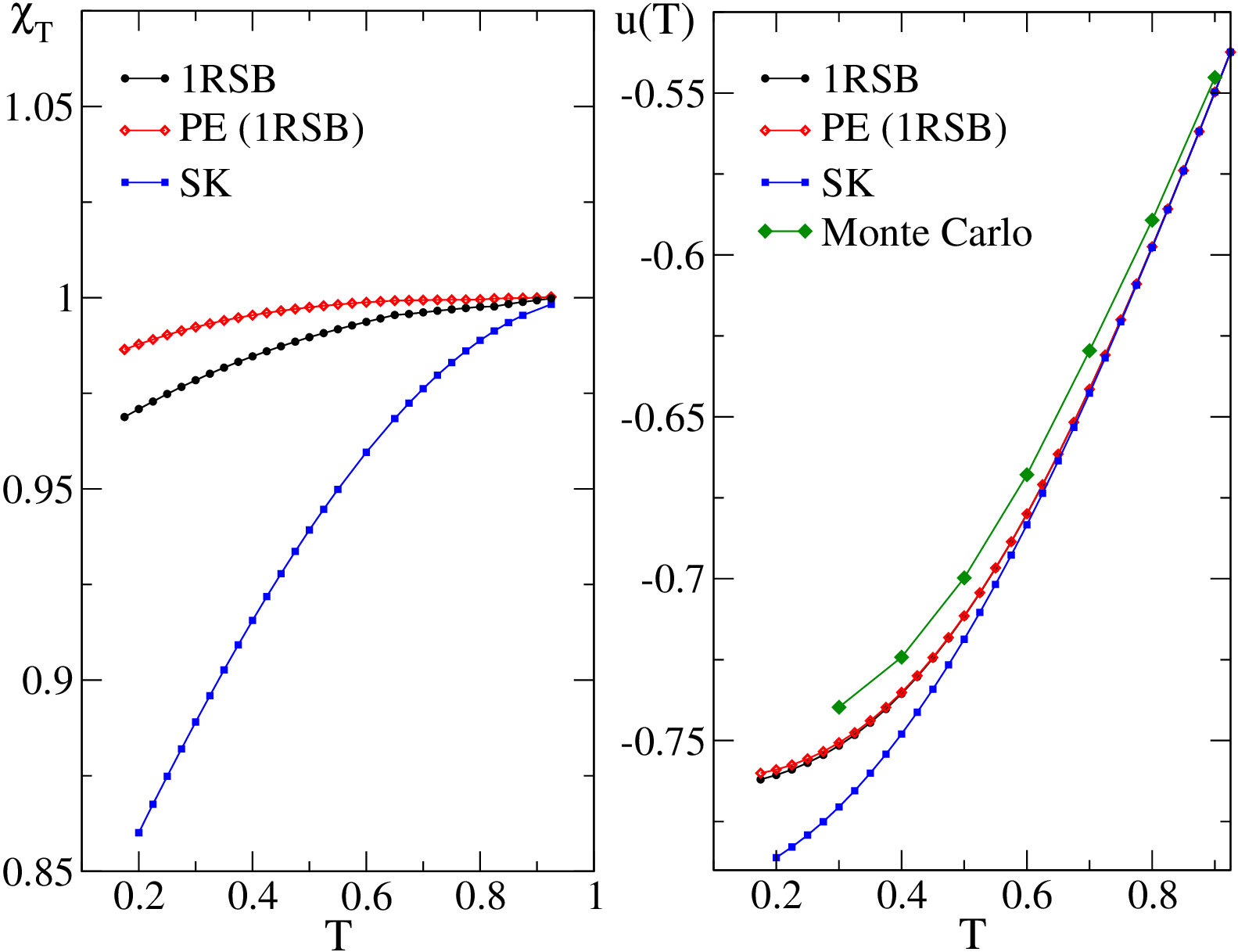}
\caption{(Color online) Temperature dependence of magnetic susceptibility $\chi_T$ (left panel) and internal energy $u$ (right panel)  calculated in 1RSB scheme, with first correction to it and the replica symmetric solution. Monte Carlo result  for the internal energy is shown for comparison. }
\label{fig:susc_en}
\end{figure}

Failure of the SK solution to produce physical results at low temperatures can be demonstrated on entropy.  In Fig.~\ref{fig:entr_ff} we plotted entropy (left panel) and free energy (right panel) calculated in the SK, 1RSB, and the first-order expansion around 1RSB. We can see that even though there is not a big difference in free energy between the 1RSB and the expansion around it, an improvement in the entropy at low temperatures is tangible and goes beyond the high-order expansion of Ref.~[\onlinecite{Crisanti02}]. Improvement of the perturbation expansion upon the 1RSB solution at low temperatures can be demonstrated on magnetic susceptibility $\chi_{T}$ from Eq.~\eqref{eq:susceptibility} plotted in Fig.~\ref{fig:susc_en}, left panel. It is expected to stay fixed at the value $\chi_{T}= 1$ in the whole low-temperature phase.\cite{Sompolinsky81} We also plotted temperature dependence of the internal energy $u = f- TS$ and compared the result with Monte-Carlo simulations (right panel).  

\begin{figure}
\includegraphics[width=0.48\textwidth]{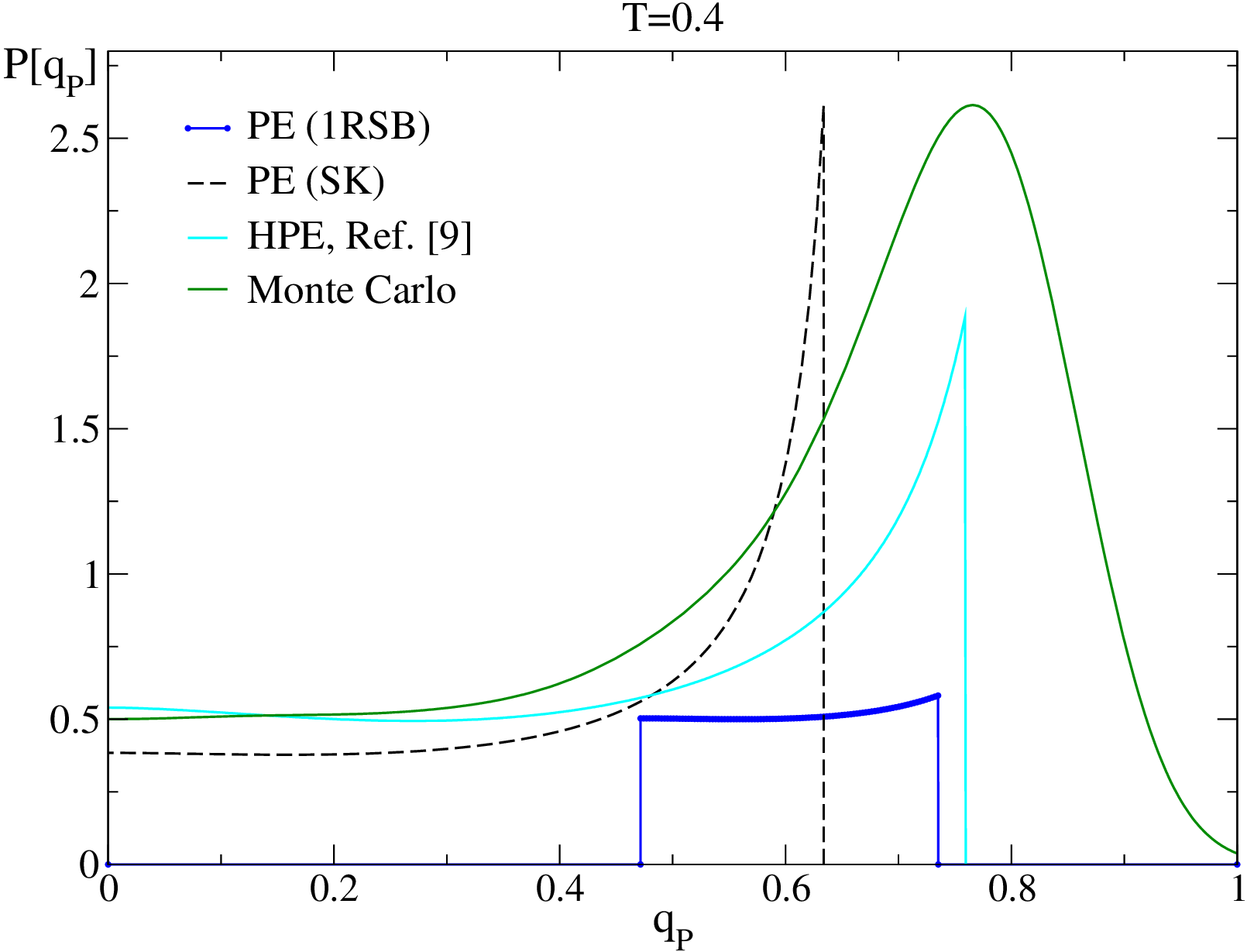}
\caption{(Color online) Derivative of the Parisi order parameter $dx_P/dq_P=P(q_P)$  corresponding in our scheme to $-\beta^{-1}\dot{\mu}(x)$ is plotted as a function of $q_{P}$ in the Parisi notation for $T=0.4$. We  also added curves from the Monte Carlo calculation and from Ref~[\onlinecite{Crisanti02}].}
\label{fig:pq_T04}
\end{figure}
\begin{figure}%\vspace{10pt}
\includegraphics[width=0.48\textwidth]{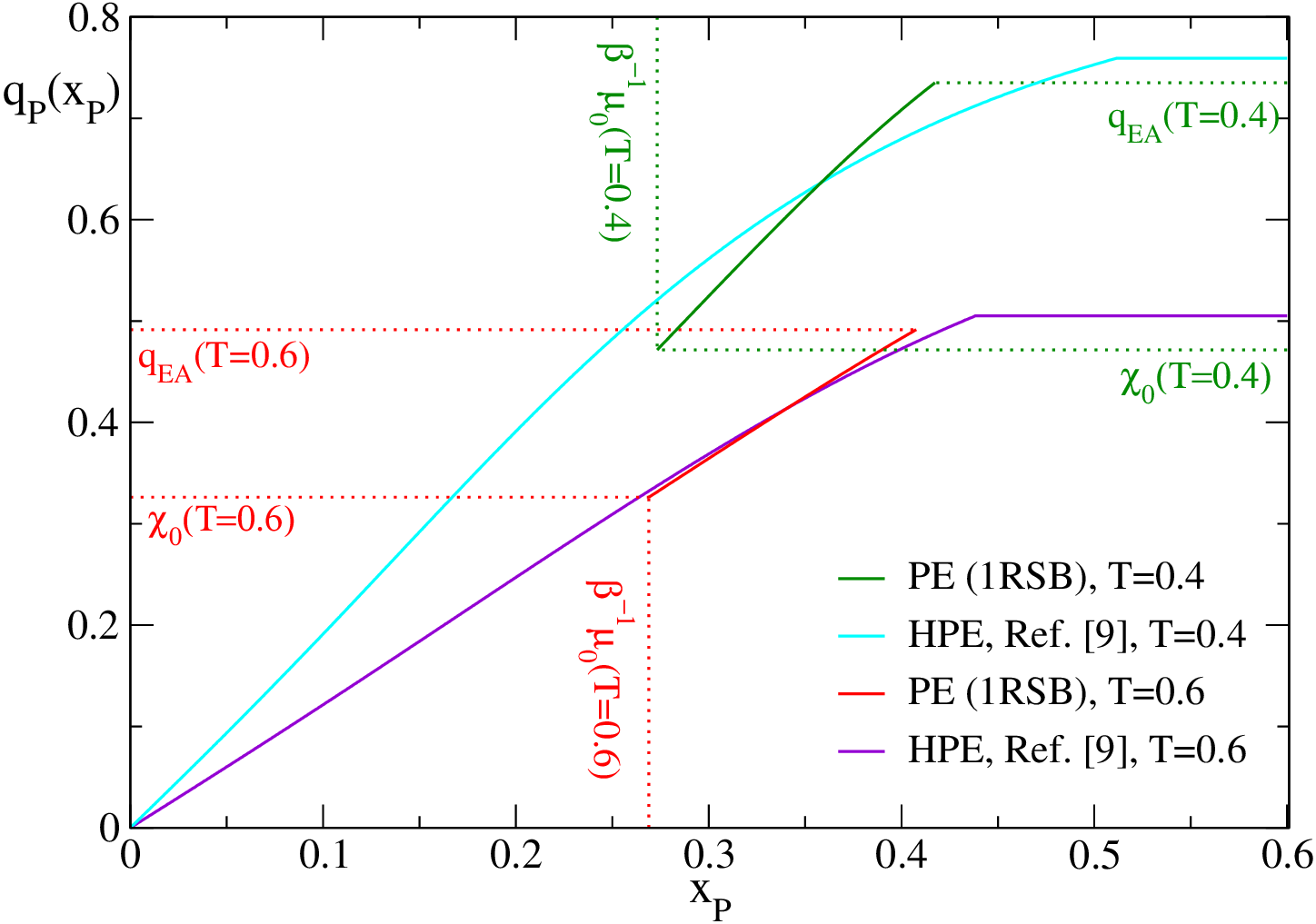}
\caption{(Color online) Parisi order parameters: $q_P(x_P)$   as obtained in Ref.~[\onlinecite{Crisanti02}] at two  different temperatures, $T=0.4$ (upper curves) and $T=0.6$ (lower curves), compared with the first order correction to 1RSB of the presented perturbation expansion.   Parameters $\chi_0$, $\beta^{-1}\mu_0$, and $q_{EA}$ are marked by dotted lines.}
\label{fig:qp_vs_xp}
\end{figure}

The continuous order-parameter function is not measurable and one cannot assess quality of approximations on this function. But according to the interpretation of the derivative $dx_{P}/dq_{P}$  in the Parisi notation,\cite{Parisi83} being in our approach $-\beta^{-1}d\mu(x)/dx$, we can compare this function with the probability distribution of overlap magnetizations $P(q) = N^{-1}\sum_{s,s'}P(s) P(s') \delta(q^{ss'}- q)$ accessible via Monte-Carlo simulations.\cite{Young83} We plotted this function at $T=0.4$ in Fig.~\ref{fig:pq_T04}. We added also the result from Ref.~[\onlinecite{Crisanti02}] and the perturbation expansion around the SK solution. 

Each of the approximations shows a maximum at $q_{P}=q_{EA}$. Except for the perturbation expansion around the SK solution, all approximations show the maximum for almost the same Edwards-Anderson parameter. The  function calculated from the expansion around 1RSB differs from the others in two aspects. First, it leads to a continuous function only between $\chi_{0}>0$ and $q_{EA}$.  Second, the slope of $P(q_{P})$ with which its maximum is reached at the upper end, $q_{EA}$, is more shallow. Function $P(q_{P})$ in the expansion around 1RSB  is not well defined at the end points, it has different limiting values from left and right. The lower end of the defining interval  for $P(q_{P})$ is  at $q_{P} = \chi_{0}\approx 0.47$.  There are indications that higher orders of the expansion around the 1RSB solution push this initial value $\chi_{0}\to 0$ at all temperatures.  Parameter $\chi_{0}$  in the 1RSB solution decreases also with increasing temperature and tends to zero when $T\to T_{c} =1$ with a constant function $P(q_{P}) \to 2$, the exact result of the asymptotic expansion around the critical temperature.\cite{Janis06b} A nonzero value of $\chi_{0}$ is, however, indispensable  in approximate treatments if we want to reach the limit to zero temperature. 

A more important feature, rather than the position of the lower bound of the definition domain of $P(q_{P})$, is the height of its step there. The expansion around 1RSB leads to $P(\chi_{0}) \approx 0.502$ and this result is not much affected by higher orders of the expansion. It reproduces well the value of Monte-Carlo simulations  with $P(0) \approx 0.50$. The results from the expansions around SK and from Ref.~[\onlinecite{Crisanti02}], being $P(0) \approx 0.38$ and  $0.54$, respectively, do not do that well.  The expansion from Ref.~[\onlinecite{Crisanti02}] overshoots the Monte-Carlo initial value and $P(q_{P})$ then decreases for low values of of $q_{P}$, displays a shallow minimum at $q_{m}\approx 0.28$ with $P(q_{m}) \approx 0.49$ before it starts to steeply grow up to reach the end point. Such a behavior is observed neither in the Monte-Carlo data and nor in the present perturbation expansion. 

The differences in the behavior of distribution function $P(q_{P})$  between the perturbation expansions from Ref.~[\onlinecite{Crisanti02}] and around 1RSB are more transparent in Fig.~\ref{fig:qp_vs_xp} where we plotted  function $q_{P}(x_{P})$ for two different temperatures, $ T= 0.6$ and $0.4$. Function $q_{P}(x_{P})$  in the latter expansion remains zero up to $\beta^{-1}\mu_{0}$  at which it jumps to $q_{P}=\chi_{0}$. It means that the expansion around 1RSB  does not allow for metastable states with averaged squared magnetization smaller than $\chi_{0}$.  Both functions almost linearly increase up to saturation, where $q_{P} = q_{EA}$.  We can see that although there is not a remarkable difference in the estimate of the Edwards-Anderson parameter, the difference between the two approaches in the value of $x_{P}$ at which $q_{P}= q_{EA}$  increases with lowering the temperature. Since the high-order perturbation expansion  from Ref.~[\onlinecite{Crisanti02}] is an asymptotic series around the critical transition temperature, the expansion around the 1RSB state contains more low-temperature data and can be considered as more reliable.

\section{Conclusions}
\label{sec:Conclusions}

Complexity and low accessibility of reliable mean-field approximations in spin-glass models lies in the continuous order-parameter function one has to introduce to reach a thermodynamically consistent solution for all temperatures. This function enters the Parisi non-linear differential equation~\eqref{eq:g0-DE} determining the equilibrium interacting free energy. This equation is unsolvable and one must resort to approximations. There is, however, no apparent way how to systematically iterate the full solution. The usual way is to use the replica trick and approximations with finite numbers of replica hierarchies. Although one improves in this way upon thermodynamic consistency, one does not learn about the actual ordering in the equilibrium state and the phase space of the order parameters.  It is hence important to have an approximate scheme addressing directly the continuous order-parameter function in the whole glassy phase, including zero temperature, where the thermodynamic inconsistency of discrete approaches is most pronounced.        
 
Here we proposed  a construction of a free energy containing the order parameters of a solution with one hierarchy of replica-symmetry breaking together with a continuous order-parameter function.  The continuous function enters the Parisi non-linear differential equation determining the interacting part of the free energy functional. We showed that to reach explicit formulas in the limit to zero temperature ($\beta \to \infty$) it is necessary to use the 1RSB free energy as the starting point for a perturbation expansion in powers of the continuous order-parameter function. The 1RSB solution not only allows for the explicit limit to zero temperature but also justifies reliability of low-order approximations to rather low temperatures, which is not the case if we apply the same expansion to the replica-symmetric solution, being equivalent to an expansion around the paramagnetic state.          

We derived explicit equations for all the variational parameters including the continuous variational function of the generating free-energy functional. We also represented relevant physical quantities without the necessity to refer to replicas and the replica trick. We showed how to construct iteratively systematic approximations to all quantities by expanding them in powers of the continuous-order parameter function beyond the 1RSB representations. We presented a closed form of an approximation containing first corrections in the expansion parameter to all quantities of interest. Comparison with Monte-Carlo simulations and the high-order asymptotic expansion of Ref.~[\onlinecite{Crisanti02}] for the Sherrington-Kirkpatrick model at zero magnetic field proves that even the lowest approximation produces reliable results and offers a qualitative picture of the behavior of the continuous order-parameter function in the whole glassy phase.       

There are two promising directions of the application of the construction developed here. First, one can investigate the zero-temperature properties and the behavior of physical quantities in the asymptotic limit $T\to 0$ such as entropy and magnetic susceptibility. Second, one can use this approximate scheme in generalized spin-glass models where one expects first-order phase transitions  with regions where a 1RSB solution is locally stable and coexists with another one with continuous replica-symmetry breaking. There are presently no approaches allowing to study quantitatively such situations.

Last but not least, we found an explicit free energy with a continuous order-parameter function that is fully thermodynamically consistent and exact to first two orders of this function. It may replace the Parisi solution for which there is no explicit or closed-form representation. Free energy~\eqref{eq:f2} is stationary in the equilibrium state with respect to all its variational parameters, including the order-parameter function $\mu(x)$. Unlike representation~\eqref{eq:FE-continuous} generating the Parisi solution, free energy $f_{2}$ from Eq.~\eqref{eq:f2} determines $\mu(x)$ from a solvable linear integral equation~\eqref{eq:mu-x-0}. Such a free energy can be understood as a solvable Landau functional for the continuous order-parameter function and may shed more light on the behavior of solutions with continuous replica-symmetry breaking. It can serve as a viable improvement upon the 1RSB approximation interpolating between the states with discrete and continuous replica-symmetry breaking in mean-field spin-glass models.

\end{document}